# How hosts and pathogens choose the strengths of defense and counter-defense. A game-theoretical view

**Shalu Dwivedi[1], Ravindra Garde[1,2], Stefan Schuster[1*]**

[1]Dept. of Bioinformatics, Matthias Schleiden Institute, University of Jena, Ernst-Abbe-Platz 2, 07743 Jena, Germany

[2]Present affiliation: Braunschweig Integrated Centre of Systems Biology, Helmholtz Centre for Infection Research, 38124 Braunschweig, Germany

shalu.dwivedi@uni-jena.de ; ravindra.p.garde@gmail.com ; stefan.schu@uni-jena.de

**\*Correspondence:**
Corresponding Author
stefan.schu@uni-jena.de

**Abstract**

Host-pathogen interactions consist of an attack by the pathogen, frequently a defense by the host and possibly a counter-defense by the pathogen. Here, we present a game-theoretical approach to describing such interactions. We consider a game where the host and pathogen are players and they can choose between the strategies of defense (or counter-defense) and no response. Specifically, they may or may not produce a toxin and an enzyme degrading the toxin, respectively. We consider that the host and pathogen must also incur a cost for toxin or enzyme production. We highlight both the sequential and non-sequential versions of the game and determine the Nash equilibria. Further, we resolve a paradox occurring in that interplay. If the inactivating enzyme is very efficient, producing the toxin becomes useless, leading to the enzyme being no longer required. Then, production of the defense becomes useful again. In game theory, such situations can be described by a generalized matching pennies game. As a novel result, we find under which conditions the defense cycle leads to a steady state or to an oscillation. We obtain, for saturating dose-response kinetics and considering monotonic cost functions, 'partial (counter-)defense' strategies as pure Nash equilibria. This implies that producing a moderate amount of toxin and enzyme is the best choice.

## 1   Introduction

A typical host-pathogen interaction involves an attack by the pathogen and defense by the host. Many pathogens invest in a counter-defense, that is, in not just attacking the host but also in bypassing or neutralizing the host defenses in order to increase the efficacy of the attack (1-6). For example, reactive oxygen species are formed by immune cells of the host and superoxide dismutases are synthesized by the pathogenic fungus *Candida albicans* to neutralize the reactive oxygen species (1). Another example is phagocytosis of *C. albicans* by macrophages, which is counter-acted by the pathogen via a yeast-to-hyphae transition (7, 8). Although not being a host-pathogen interaction in the strict sense, it is worth mentioning that in the competition between *Streptomyces clavuligerus* and Salmonella bacteria, a counter-counter defense can be observed: clavulanic acid secreted by *S. clavuligerus* inhibits ß-lactamases, which inactivate ß-lactam antibiotics (9, 10). A further biological example is the defense of plants such as *Arabidopsis thaliana* against the bacterial pathogen *Pseudomonas syringae* by RNA silencing and a counter-defense by suppressors of that process (11).

Similar effects occur in plant-herbivore interactions. Many plants such as *Diplopterys cabrerana* (12), *Psychotria viridis* (12, 13), *Mimosa tenuiflora* (14) and also the fungal genus *Psilocybe* (14, 15) synthesize tryptamine compounds as defense chemicals. Higher animals produce, mainly in the brain, monoamine oxidase, which degrades tryptamines (12, 14, 15). To give another example, Brassicaceae plants produce toxic isothiocyanates (ITCs) from glucosinolates (GLSs) as soon as they are attacked by herbivores (16-18). Here, various counter-defenses are observed where specialist herbivores detoxify GLSs by preventing GLS hydrolysis to ITCs, while the generalist herbivores cannot block GLS hydrolysis but detoxify the ITCs to a small extent. As a result, the generalists suffer on Brassicaceae hosts (16-18).

To understand host-pathogen or plant-herbivore interactions, mathematical modeling and computer simulation have turned out to be very helpful (19-24). This can be done, among other methods, by systems of ordinary differential equations (19, 25) or by evolutionary game theory (4, 8, 23, 26-35). Previously, a mathematical model based on enzyme-kinetic equations describing the interplay between a defense chemical, enzymes degrading those, and inhibitors of the enzyme was proposed (10). The calculations show that only in the case of strong binding of the inhibitor, it pays to produce an inhibitor as a counter-counter defense.

However, a paradox occurs in the above-mentioned interaction. If the counter-defense (e.g. inactivating enzyme) is very efficient, the defense (e.g. toxin) becomes useless, so that there is no longer any need for its production. Then, the counter-defense becomes unnecessary. If, however, this is stopped being produced as well, production of the defense becomes useful again (22). The question arises whether this leads to an oscillatory change in strategies or instead to a steady state being a trade-off, in which the two species produce a moderate amount of toxin and enzyme, respectively. Although this question is very relevant for medical and pharmacological applications, it has hardly been analyzed so far for such applications. In game theory, such situations without a clear equilibrium are known as (generalized) "matching pennies game" (36, 37). In its basic, traditional version, each of two players secretly turns a penny to heads or tails and then they show it at the same time. If the pennies match, that is both heads up or both tails up, then player 1 keeps both pennies. If the pennies do not match, player 2 keeps both pennies (37). Both players have the same set of strategies and only the payoffs 1 and −1 occur. Nevertheless, it is an asymmetric game. In the generalized version, more than two different payoff values occur and the game is not usually a zero-sum game anymore (36).

The "generalized matching pennies game" can be found in several situations in biology. For example, when a prey can hide in either of two different locations and a predator animal only has the time or energy to search in one of the locations (38). A related situation is when a predator (e.g. a leopard) chases a prey (e.g. a gazelle), the predator would try to turn to the same side as the prey is evading, while the prey tries to avoid that (39). Also with respect to the decision between daytime activity and nighttime activity of predator and prey, a matching pennies game can be observed. Applications other than in biology include auditing in management and penalty shooting in soccer (36, 39). The goal keeper has a strong motivation to jump to same side as the ball is coming, while the kicker prefers a mismatch. Similarly, the prohibition of alcohol in the U.S. in the 1920's can be regarded from that viewpoint. That law finally became useless because of clandestine brewing and distilling.

The correspondence to the problem studied here can be seen by relating "matching" (head/head or tail/tail) to producing (counter-)defense by both sides or by neither side. Then, the pathogen has an advantage. If the strategies do not match (defense without counter-defense or vice versa), the host has an advantage because it either can defend itself or the pathogen bears the cost of counter-defense for nothing.

Here, we present a game-theoretical analysis to answer the above questions in an innovative way. We consider both the sequential and non-sequential version of the game. The analysis considerably extends a preliminary game-theoretical interpretation presented earlier (22). The host and pathogen are considered as players and they maximize their payoffs (outcome, which can be quantified as gain in fitness) by choosing appropriate strategies.

As has been done in several other game-theoretical studies on host-pathogen interactions (4, 27-32), we here use classical game theory, which revolves around the concepts of payoff matrix and Nash equilibrium. The latter refers to an equilibrium solution of the game in which neither player has an incentive to change strategy unilaterally (35, 37). A more sophisticated approach (often called Evolutionary game theory) is based on the concept of evolutionary stable strategy, which allows one to study the fate of a rare mutant with a new strategy within a resident population (35, 40, 41). The classical approach used here provides a basic understanding of the interactions under study (27, 28). It is worth mentioning that also in classical game theory, the possibility is taken into account that a population subdivides into subpopulations adopting different strategies and that they either coexist or that one subpopulation outcompetes the other one.

Although only rarely discussed in game theory, it is tacitly assumed that the Nash equilibrium in non-sequential games is often found by iteration, in particular if there is more than one equilibrium. One player is choosing a preliminary strategy, the other player is responding and then the first player may change strategy according to that response, and so on until an equilibrium is found.

An important point is whether the players are allowed to see what the other one is doing, that is, whether information exchange is allowed. Here, we assume that players know each other's strategies and can choose their strategy accordingly. This is justified because organisms can usually (but not in every case) sense effector molecules produced by other organisms (42, 43). Micro-organisms can obtain information about their surroundings (44). It may be speculated that some microbes can even anticipate what the other player will do. In any case, an interpretation in terms of populations can be used. Consider a population of hosts including variants (e.g. mutants) that choose one strategy and other variants choosing the other strategy. Likewise, there is a population of pathogens including two variants. When all of them interact in the sequential game, four different outcomes will occur. Now, the variants getting a higher payoff will win the competition against the variants getting a lower one.

To make the presentation clearer, we here consider the situation where the pathogen produces an enzyme (such as a superoxide dismutase) that partially or completely degrades or inactivates a defense chemical (such as reactive oxygen species) produced by the host. However, the analysis can be applied to many other biological 'Defense/ Counter-defense' systems.

The assumptions and simplifications considered here in the game settings can be summarized as follows:

- Asymmetric game among a population of hosts and a population of pathogens
- Each individual (player) has two strategies (in Section 2), notably (counter-)defense or no (counter-)defense, or three strategies (in Subsection 3.2), which include partial (counter-)defense

- Individuals within one and the same population may use different strategies
- Individuals getting a higher payoff will win the competition within a population
- The solution of the game is found by random events in populations

Further assumptions are mentioned for the different scenarios in the respective subsections.

## 2 Modeling the case where each player has two strategies

Obviously, both the host and pathogen have a broad spectrum of possible strategies in that they can vary the levels of toxins and toxin-degrading enzymes. First we consider the simplest case where the host only has two strategies,

1) D: Defense

2) ND: No defense

and the pathogen only has two strategies,

1) CD: Counter-defense

2) NCD: No counter-defense.

This is an asymmetric game because the sets of strategies and also the payoff values differ for the two players (Figure 1). We distinguish between non-sequential (simultaneous) and sequential games depending on whether the choice is made simultaneously (or by iteration, see Introduction) or, alternatively, the two players choose their strategies after each other and cannot change them afterwards. In the first case, we use payoff matrices, while in the second case, we use both (extended) payoff matrices and the method of game trees (35).

Sometimes, the enzyme is not able to degrade the toxin completely. We categorize the strength of enzyme as *perfectly efficient counter-defense* and *imperfectly efficient counter-defense*.

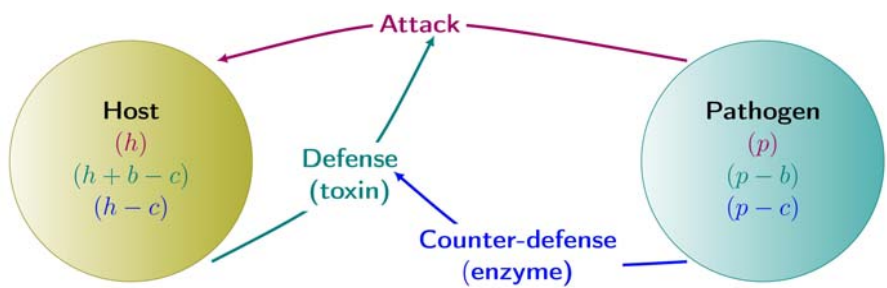

**Figure 1. Schematic diagram for the defence and counter-defense game in host-pathogen interactions.** $h$, $p$, initial payoffs of host and pathogen, respectively; $b$, benefit from using toxin; $c$, cost of producing defense and counter-defense.

## 2.1 Perfectly efficient counter-defense

First, we assume that the enzyme completely inactivates the toxin. Now, as mentioned above, we consider two scenarios according to the sequence in time.

### 2.1.1 Non-sequential game

First, we consider the non-sequential or simultaneous game. The payoff matrix (also known as normal form) for this subcase is given in Table 1 (see also Figure 1).

**Table 1:** Payoff matrix for host-pathogen interaction (perfectly efficient counter-defense).*

| Host | Pathogen | |
|---|---|---|
| | No counter-defense (NCD) | Counter-defense (CD) |
| No defense (ND) | $(h, p)$ ↓ | ← $(h, p - c)$ |
| Defense (D) | $(h + b - c, p - b)$ → | ↑ $(h - c, p - c)$ |

*No pure Nash equilibrium occurs for the high-benefit case ($b > c$). A path following incentives of the two players leads to a cycle, as indicated by arrows. The equilibrium is 'ND/NCD' for the low-benefit case ($b < c$) (pink).

Here, we denote the payoff value of the host and pathogen when the latter attacks the former, yet without any defense and counter-defense, as $(h, p)$, the cost of defense or counter-defense as '$c$' and the respective benefit as '$b$'. For simplicity's sake, we assume that the costs are the same for both organisms and that a benefit for the host implies an opposite effect for the pathogen. This restriction can be relaxed to some extent, as long as the order relations between the payoffs assumed in the two cases considered below are not changed.

We first assume that the counter-defense (e.g. inactivating enzyme) is highly efficient, so that the defense is practically useless. So the payoff values in the case D/CD are just given by the payoffs in the case ND/NCD minus the costs, i.e. $(h - c, p - c)$. If the pathogen chooses to counter-defend against no response by the host, then only the pathogen must pay the cost without having any gain. So the payoff value in the case ND/CD is $(h, p - c)$. If the host chooses to defend against no response by the pathogen, then only the host must pay the cost. In this case, the host and pathogen get benefit ($b$) and harm ($-b$). So the payoff value of D/NCD is $(h + b - c, p - b)$.

We distinguish two cases, which differ in the order relations between the payoffs.

*(i) High-benefit case: $b > c$*

In the payoff matrix (Table 1), we can start from ND/NCD i.e. $(h, p)$. Here, host has an incentive to change its strategy to D as $h + b - c > h$. Now, the pathogen has an incentive to change its strategy to CD as $p - c > p - b$. This, however, incites the host to switch its strategy to ND again and save its cost because $h > h - c$. Here, the pathogen has again an incentive to switch its strategy to NCD as $p > p - c$. This gives host a choice to produce defense against the pathogen since $h + b - c > h$ and so on. This leads to oscillations. Starting from another cell in the matrix does not change this result because the oscillation covers all cells. So, no pure Nash equilibrium occurs.

The absence of a pure Nash equilibrium is characteristic for many games studied in game theory earlier (32-35). For example, in the "matching pennies" game, each of two players secretly turns a penny to heads or tails and then they show it at the same time. If the pennies match, that is both heads up or both tails up, then player 1 keeps both pennies. If the pennies do not match, player 2 keeps both pennies (35). The matching pennies game has a more symmetric structure than the defense/counter-defense game studied here because both players have the same strategies. However, it is an asymmetric game because the payoffs are not the same for both players. The correspondence between the two games can be seen by relating "matching" (head/head or tail/tail) to producing (counter-)defense by both sides or by neither side. Then, the pathogen wins. If the strategies do not match (defense without counter-defense or vice versa), the host has an advantage because it either can defend itself or the pathogen bears the cost of counter-defense for nothing.

In the absence of a pure Nash equilibrium, still a mixed Nash equilibrium occurs, which implies in the two-player game that each player chooses one or the other strategy with certain probabilities (35, 40, 45). This can be interpreted in biological terms in different ways. Either, in accordance with the cyclic dominance in the payoff matrix, oscillations occur in that defense and no defense alternate on evolutionary time scales. Moreover, there is a natural correspondence between the mixed Nash equilibria of a two-player normal form game and the Nash equilibria of an asymmetric evolutionary game in populations [(35), Section 9.15]. In the game between populations, the probabilities are usually interpreted as fractions of the populations adopting a certain strategy (35, 40). In populations, oscillations can be interpreted in that the relative frequencies of the different strategies oscillate in time around the equilibrium frequencies, as often discussed in the case of the famous rock-scissors-paper game, which also shows a mixed Nash equilibrium (35, 46, 47). For the game under study here, defense and no defense may alternate on evolutionary time scales. A second interpretation is that some hosts use a defense while others do not, and they stay with their strategy. A third option is that all individuals within one population permanently produce a certain fraction of the maximum amount of (counter-)defense possible. Organisms may have a broader spectrum of possible actions in that they can vary the concentrations of toxins and toxin-degrading enzymes. In order to still consider a discrete set of strategies, a straightforward extension is to consider three rather than two strategies. We will analyze these three options in Section 3.

Let us denote the fraction of hosts choosing defense or no defense by $f_D$ and $f_{ND}$, respectively, and analogously $g_{CD}$ and $g_{NCD}$ for the pathogen population. We assume that the hosts use the 'defense' strategy, with probability $f_D$ and 'no defense' with probability $1-f_D$, i.e. $f_{ND}$. This could be interpreted in that a fraction $f_D$ of all hosts use the defense strategy. Similarly, the pathogens use the 'counter-defense' strategy with probability $g_{CD}$ and 'No counter-defense' with probability $1-g_{CD}$, i.e. $g_{NCD}$.

For the system under study, the expected payoff, $P_H$, of the hosts read,

$$P_H(f_D, g_{CD}) = h(1 - f_D)(1 - g_{CD}) + h(1 - f_D)g_{CD} + (h + b - c)f_D(1 - g_{CD}) + (h - c)f_D g_{CD}$$

$$(b - c - bg_{CD})f_D + h.$$

Similarly, the expected payoff, $P_p$, of the pathogens read:

$$P_P(f_D, g_{CD}) = p(1 - f_D)(1 - g_{CD}) + (p - c)(1 - f_D)g_{CD} + (p - b)f_D(1 - g_{CD}) + (p - c)f_D g_{CD}$$

$$(bf_D - c)g_{CD} - bf_D + p$$

The equilibrium fractions of the host can be calculated by setting the derivatives of the average payoffs of the host w.r.t. the fractions corresponding to that player equal to zero (35, 40):

$$\frac{\partial P_H}{\partial f_D} = 0 \Rightarrow (b - c - bg_{CD}) = 0 \text{ gives } g_{CD} = \frac{b-c}{b}$$

Analogously, the equilibrium fractions of the pathogen are calculated. Together, this gives:

$$\frac{\partial P_P}{\partial g_{CD}} = 0 \Rightarrow (bf_D - c) = 0, \text{ gives } f_D = \frac{c}{b}$$

$$f_D = \frac{c}{b}, f_{ND} = \frac{b-c}{b}, g_{CD} = \frac{b-c}{b}, g_{NCD} = \frac{c}{b} \tag{1a,b,c,d}$$

Since in the high-benefit case, $b > c$, all these values are between 0 and 1.

*(ii) Low-benefit case: $b < c$*

We can start from ND/NCD, i.e. $(h, p)$ (Table 1). The host gets less payoff if it produces defense against the pathogen as the cost is higher than the benefit, $h + b - c < h$. So strategy D is not in favor of the host. While the pathogen must pay the cost without any gain if it chooses CD against the host which reduces its payoff i.e. $p - c < p$. They do not have any incentive to change their strategies. Hence, the Nash equilibrium is 'ND/NCD'. This is understandable because the costs for defense or counter-defense exceed the corresponding benefit.

Alternatively, we may start from another cell in the matrix, for example, ND/CD, i.e. $(h, p - c)$. Here, strategy D is not in favour of the host as $h - c < h$ while the pathogen can switch its strategy to NCD and save its cost. Hence, the Nash equilibrium is again 'ND/NCD'. We reach the same equilibrium when starting from any other cell.

### 2.1.2 Sequential game

In the sequential game the second player can play a strategy after observing the first player's move. In a sequential interaction, it is biologically plausible that the host "decides" first whether or not to produce a toxin and only subsequently, the parasite "decides". (The fact that it is the pathogen that initiates the interaction by attacking the host is neglected here because we do not consider that level.) Out of academic interest, we also consider, in the Supplement, the opposite case where the parasite decides first.

The host can choose either to defend or not to defend, while the pathogen has now four strategies:

i) No counter-defense no matter what host does (NCD)

ii) Counter-defense no matter what host does (CD)

iii) Do the same thing as host does (NCD/CD)

iv) Do the opposite of what host does (CD/NCD)

We can write the payoff matrix on the basis of two and four strategies for the host and pathogen, respectively (Table 2).

**Table 2:** Payoff matrix if host plays first (perfectly efficient counter-defense).*

| Host | Pathogen | | | |
|---|---|---|---|---|
| | i) NCD | ii) CD | iii) NCD/CD | iv) CD/NCD |
| ND | $(h, p)$ | $(h, p-c)$ | $(h, p)$ | $(h, p-c)$ |
| D | $(h+b-c, p-b)$ | $(h-c, p-c)$ | $(h-c, p-c)$ | $(h+b-c, p-b)$ |

*The Nash equilibrium is 'ND/NCD' for both the high-benefit case ($b > c$) (yellow) and low-benefit case ($b < c$) (pink).

An alternative representation is by using game trees, also known as extensive form game (35) (Figure 2). Root and branch nodes represent players and branches represent their strategies. Leaves or terminal nodes correspond to outcomes of the game with the respective payoffs. A convenient method for finding the Nash equilibria in game trees is by eliminating the dominated strategies from the leaves to the root, a method called backward induction.

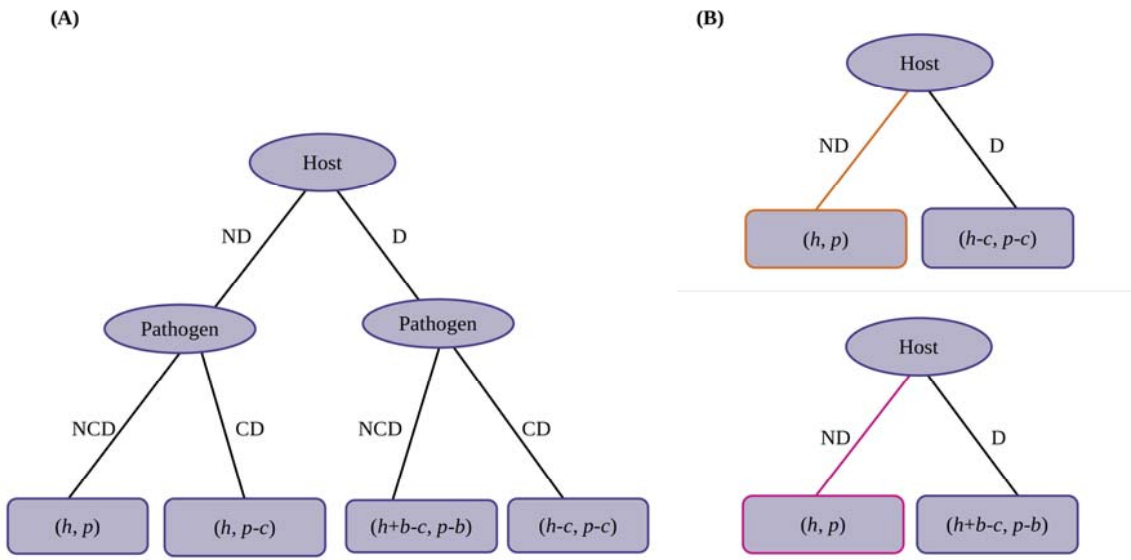

**Figure 2. Game tree (extensive form game) for the case where the counter-defense is perfectly efficient.** A) Initial tree. B) Processed form. Top panel, case (i) ($b > c$). Bottom panel, case (ii) ($b < c$). Nash equilibria are indicated in yellow and pink corresponding to Table 2.

We distinguish two cases, which differ in the order relations between the payoffs.

*(i) High-benefit case: b > c*

Here, we use backward induction to solve the game tree. We start from the left leaf nodes $(h, p)$ and $(h, p - c)$ (Figure 2A). Strategy NCD dominates CD as the payoffs for the pathogen satisfy the inequality $p > p - c$. So, we eliminate strategy CD. Now, in the right nodes $(h + b - c, p - b)$ and $(h - c, p - c)$, strategy CD dominates NCD as $p - c > p - b$. So we eliminate strategy NCD.

The next processed tree is shown in Figure 2B (top panel). Here, strategy ND dominates D as $h > h - c$. So, the host will choose strategy ND. The final payoffs for the host and pathogen are '$h$' and '$p$'. Hence, the equilibrium is 'ND/NCD'. Intuitively, one may have not expected this result if benefits are high.

*(i) Low-benefit case: b < c*

In this case (Figure 2A), we start from the left and right leaf nodes as before and we see that strategy NCD dominates CD on both sides as the payoffs for the pathogen satisfy the inequality $p > p - c$ and $p - b > p - c$. So, we eliminate strategies CD from both sides.

In the tree, shown in Figure 2B (bottom panel), strategy ND dominates D as inequality $h > h + b - c$. The final payoffs are '$h$' and '$p$'. Hence, the equilibrium is 'ND/NCD'.

Interestingly, the Nash equilibrium is the same for both cases (i) and (ii), i.e. 'ND/NCD'. Obviously, this is not observed in all real interactions between a host and a pathogen. In the Introduction, we gave several examples of chemical defenses. This discrepancy may arise from our simplifying assumption that the counter-defense is perfectly efficient.

## 2.2 Imperfectly efficient counter-defense

Now we assume that the counter-defense is not perfectly efficient. This means, for example, that the enzyme degrades the toxin partially, so that there remains some effect of the toxin after defense and counter-defense.

### 2.2.1 Non-sequential game

The payoff matrix for the non-sequential game where the host and pathogen choose their strategies simultaneously is given in Table 3.

**Table 3:** Payoff matrix for host-pathogen interaction (imperfectly efficient counter-defense).*

| Host | Pathogen | |
|---|---|---|
| | No counter-defense (NCD) | Counter-defense (CD) |
| No defense (ND) | $(h, p)$ | $(h, p - c)$ |
| Defense (D) | $(h + b - c, p - b)$ | $(h + b/2 - c, p - b/2 - c)$ |

*The Nash equilibria are 'D/CD' for the high-benefit case ($b/2 > c$) (yellow), 'D/NCD' for the intermediate-benefit case ($b/2 < c < b$) (purple) and 'ND/NCD' for the low-benefit case ($b < c$) (pink).

Since we assume that the counter-defense (e.g. inactivating enzyme) is imperfectly efficient, the defense is not completely useless. So the payoff values in the case D/CD are just given by the payoffs $(h, p)$ modified by the costs and some benefit/loss due to the toxin, which we here assume to be half of the full benefit, leading to the payoffs $(h + b/2 - c, p - b/2 - c)$. All other payoffs are the same as before. In the scenarios of imperfectly efficient counter-defense, we distinguish three subcases:

*(i) High-benefit case: b > c*

We start from the cell ND/NCD. Here, the host has an incentive to change its strategy to D due to the inequality $h + b - c > h$. Now, the pathogen changes its strategy to CD as $p - b/2 - c > p - b$ is satisfied. At this point, they do not have any incentive to change their strategies. In a similar way, starting from every cell of the payoff matrix, we find that the iteration stops at the cell D/CD every time. Hence, the equilibrium is 'D/CD'.

*(ii) Intermediate-benefit case: b/2 < c < b*

In this case, the equilibrium is 'D/NCD' with the payoffs $(h + b - c, p - b)$ because neither player then has an incentive to change its strategy due to the inequalities defining this case. The other entries in the matrix do not have this property, so that this is the only Nash equilibrium.

*(iii) Low-benefit case: b < c*

Here, the equilibrium is 'ND/NCD'. This can be understood because costs for defense and counter-defense are high.

### 2.2.2 Sequential game

Now, we consider the sequential game where the host and pathogen play as first and second player, respectively. The opposite case where the pathogen decides first is considered in the Supplement.

The host has two strategies, while the pathogen has four strategies (Table 4), notably the same as in the case of perfectly efficient counter-defense.

**Table 4:** Payoff matrix if host plays first (imperfectly efficient counter-defense).*

| Host | Pathogen | | | |
|---|---|---|---|---|
| | i) NCD | ii) CD | iii) NCD/CD | iv) CD/NCD |
| ND | $(h, p)$ | $(h, p - c)$ | $(h, p)$ | $(h, p - c)$ |
| D | $(h + b - c, p - b)$ | $(h + b/2 - c, p - b/2 - c)$ | $(h + b/2 - c, p - b/2 - c)$ | $(h + b - c, p - b)$ |

* The Nash equilibria are 'D/CD' for the high-benefit case ($b/2 > c$) (yellow), 'D/NCD' for the intermediate-benefit case ($b/2 < c < b$) (purple) and 'ND/NCD' for the low-benefit case ($b < c$) (pink).

The game-tree form of the case where host plays first and the corresponding payoffs are depicted in Figure 3.

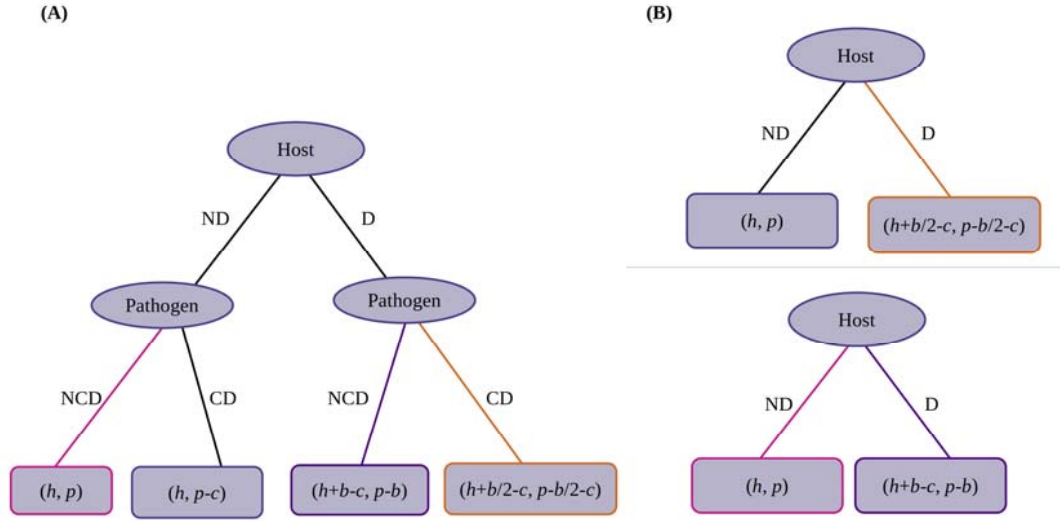

**Figure 3. Game tree (extensive form game) for the case where the counter-defense is imperfectly efficient.** A) Initial tree. B) Processed form. Top panel, case (i) (*b/2 > c*). Bottom panel, case (ii) (*b/2 < c < b*) and case (iii) (*c > b*). Nash equilibria are indicated in yellow, purple, and pink corresponding to Table 3.

Here, we again solve the game tree using backward induction. We start from the left leaf nodes and see that the strategy NCD dominates CD for the pathogen (Figure 3A). For the right leaf nodes, payoffs depend on '*b*' and '*c*'.

*(i) High-benefit case: b > c*

In the right nodes, strategy CD dominates NCD as $p - b/2 - c > p - b$. Now in the processed form, strategy D dominates ND as $h + b/2 - c > h$ (Figure 3B, top panel). So, the host will choose to defend. Hence, the equilibrium is 'D/CD'.

*(ii) Intermediate-benefit case: b/2 < c < b*

In an analogous way as above, we derive the processed tree as shown in Figure 3B (bottom panel). Here, strategy D dominates ND as $h + b - c > h$. Hence, the equilibrium is 'D/NCD'.

*(iii) Low-benefit case: b < c*

By analysing the payoffs, one can see that the next processed tree is the same as in case (ii), shown in Figure 3B (bottom panel). Here, however, strategy ND dominates D as it gives a higher payoff to the host, i.e. $h > h + b - c$. Hence, the equilibrium is 'ND/NCD'. Thus, in the three cases, three different Nash equilibria are obtained.

## 3 Modeling the case where each player has more than two strategies

### 3.1 Continuous description

In the previous section, we made a distinction between two discrete strategies for each player: (counter-)defense or no (counter-)defense. However, organisms have a broader spectrum of possible actions in that they can vary the concentrations of toxins and toxin-degrading enzymes, so that they could find a compromise equilibrium in which either side produces a submaximal amount. As mentioned in Subsection 2.1.1, we want to find out, by looking at the strategies in more detail, whether the mixed Nash equilibrium leads to such a stationary compromise or to a never-ending switching between production and no production. In this section, we calculate the payoffs for the host and pathogen from the concentrations of the toxin and enzyme, based on the benefit, which we will here call response (in the sense of dose-response curves) and the cost.

Payoff = response − cost  (2)

As we can assume that the cost for producing the toxin or enzyme is proportional to their initial concentrations (48), we have

$Payoff\ (host) = response - a * T_0$  (3)

$Payoff\ (pathogen) = -f * response - c * E_0$  (4)

where, $a$ = specific cost for producing toxin, $f$ = coefficient of toxin effect on pathogen, $c$ = specific cost for producing enzyme, $T_0$ = initial toxin concentration and $E_0$ = initial enzyme concentration.

We calculate the response from the Hill equation, which is often used for quantifying dose-response relationships (49, 50), i.e.

$$response = l + \frac{(n-l)T_0^k}{m^k + T_0^k}$$  (5)

where, $l$ = response when dosage $T_0$ is 0, $n$ = response for an infinite dosage, $m$ = half-saturation constant and $k$ = Hill coefficient

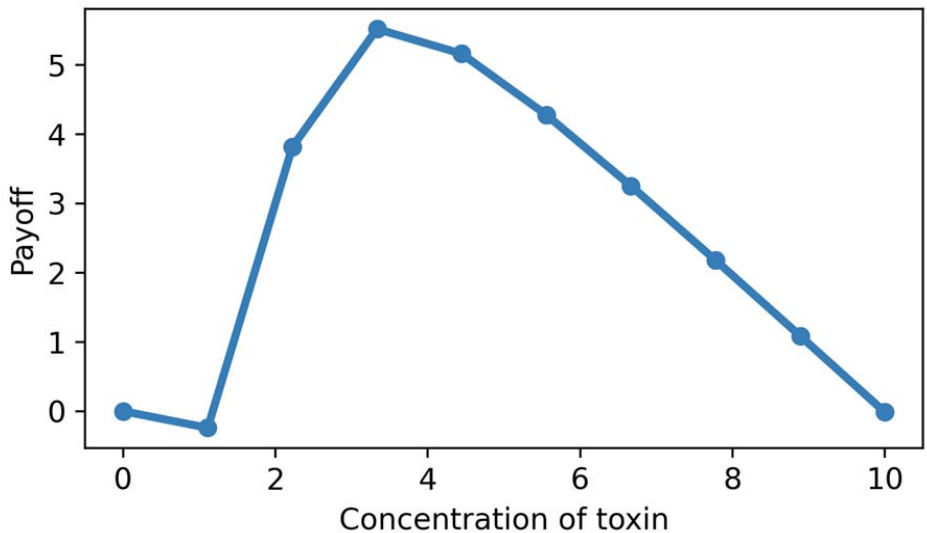

**Figure 4. Dose-payoff curve for the host (Hill equation minus linear cost function).** Parameter values: $l = 0$, $m = 2$, $n = 10$, $k = 4$.

For simplicity's sake, we put the parameter $l$ equal to zero since there is often no effect in the absence of toxin. Then, the parameter $m$ equals the value of $T_0$ at which half of maximum response occurs. Figure 4 represents the payoff as a function of toxin concentration, where the curve first decreases to negative values because for low concentrations, the toxic effect is negligible while some cost needs to be afforded. Only at higher concentrations, the curve starts increasing and reaches its maximum point and then monotonically decreases. It is worth mentioning that the graphs in Figures 4-6 were plotted with arbitrary parameters. Therefore, the numbers in the table are not based on experimental data, since these are not needed to show the qualitative effect.

A graphical model of the dependence of benefit on the allocation to defense including the effect of costs was proposed earlier by Simms and Rausher (51). It combines a Michaelis-Menten curve with linear costs and also leads to the phenomenon that the maximum effect is reached at intermediate toxin levels. A refined model based on a sigmoidal function has been suggested by Siemens and coworkers (24).

Then, we calculate the payoffs for the host from eq. (3) in the presence of the enzyme. It is not straightforward how the enzyme is taken into account. In principle, a small constant enzyme concentration is sufficient to inactivate a large toxin concentration in the long run. Here, however, we consider a snapshot, so to speak. That is, we analyze the short-term effect. We describe that effect approximately by dividing the Hill kinetics by the term $(1 + E_0)$. It has the favorable properties that it does not change the Hill kinetics in the case $E_0 = 0$ and tends to zero for very large values of $E_0$. A similar dependence was used for quantifying the effect of defense on herbivory (48). From eqs. (3) and (5), we obtain

$$Payoff\ (host) = \frac{1}{1+E_0}\left(\frac{nT_0^k}{m^k+T_0^k}\right) - aT_0 , \qquad (6)$$

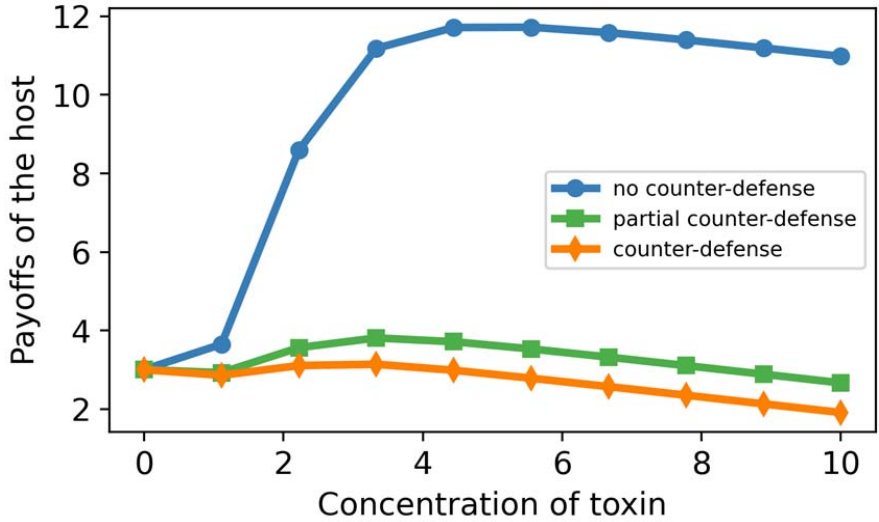

**Figure 5. Dose-payoff curve for the host as a function of toxin concentration for three different enzyme concentrations.** Parameter values: $m = 2$, $n = 10$, $k = 4$, $a = 0.2$, $E_0 = 0$ (blue curve), 5 (green) and 10 (orange).

To calculate the maximum payoff, we equate the derivative of the payoff with respect to toxin to zero. We obtain a polynomial equation for $T_0$, which cannot normally be solved analytically:

$$T_0^k - \sqrt{\frac{nkm^k}{a(1+E_0)}} T_0^{\frac{k-1}{2}} + m^k = 0$$

For the game-theoretical analysis, we need not calculate the optimum value (see below).

Figure 5 represents the payoffs of the host in dependence on the extent of defense. We can see that the graphs first decrease to negative values and then start increasing to their maximum points and then monotonically decrease. The three curves represent the payoffs of the host in the cases of no counter-defense, partial counter-defense and full counter-defense produced by the pathogen.

Obviously, a host has the highest payoff in the absence of pathogen's counter-defense. In the presence of toxin-degrading enzymes, the host has maximum payoff when it uses a sub-maximum concentration of toxin because a higher concentration would imply unnecessarily high costs (Figure 5).

Further, we calculate the payoff for the pathogen based on the toxin concentration from eq. (4), using the Hill equation in the presence of the enzyme, from eq. (5),

$$\textit{Payoff (pathogen)} = \frac{-f*response}{(1+E_0)} - cE_0 \;, \tag{7}$$

We equate the derivative of the payoff with respect to enzyme to zero. We obtain

$E_0 = \sqrt{\frac{f*response}{c}} - 1$, where the response is described in eq. (5).

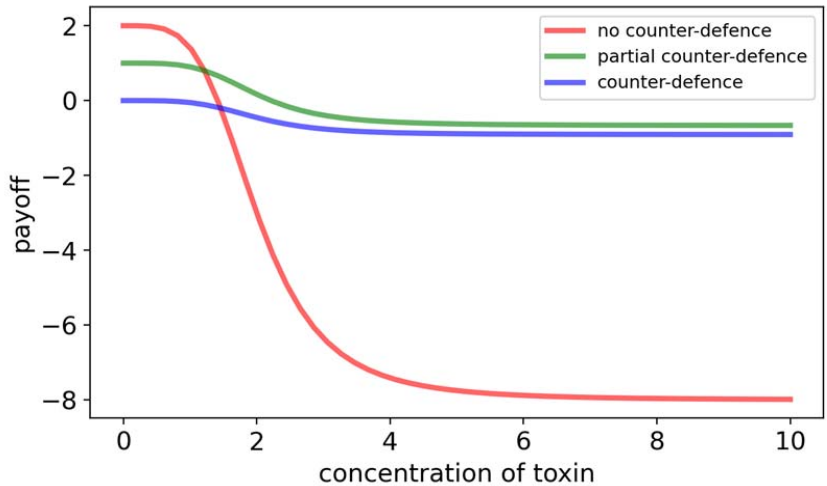

**Figure 6. Dose-payoff curve for the pathogen as a function of toxin concentration for three different enzyme concentrations.** Parameter values: $m = 2$, $n = 10$, $k = 4$, $a = 0.2$, $f = 1$. $E_0 = 0$ (blue curve), 5 (green) and 10 (orange).

Figure 6 represents the payoffs of the pathogen in dependence on the toxin concentration when it produces no counter-defense, partial counter-defense and full counter-defense. The graphs first show a short plateau, then decrease and finally tend asymptotically to negative values. The pathogen has always maximum payoffs when there is no defense produced by the host. Further, it becomes clear from Figure 6 that the production of enzyme is beneficial for the pathogen whenever the toxin concentration is above a certain threshold. The payoffs for the partial counter-defense are always higher than the payoffs for full counter-defense. This shows that the use of partial concentration of enzyme is favorable to the pathogen.

## 3.2 Description as a three-strategy game

In game theory, it is much more usual to consider discrete strategies than continuous strategies (30, 35, 40). Most often, the analysis is based on a set of two discrete strategies, as done in Section 2. However, Figure 4 shows that the highest effect is achieved for an intermediate toxin concentration. Therefore, we now distinguish three strategies for each player: no, partial or complete (counter)defense. In another context, we have analyzed a two-player, three-strategy game earlier (47). We can extract the payoff values from Figures 5 and 6 by considering the concentration values 0, 5 and 10 both for the toxin and enzyme (Table 5).

**Table 5:** Payoff values in the three-strategy host-pathogen game (from Figures 4, 5).*

| Host | Pathogen | | |
|---|---|---|---|
| | No counter-defense | Partial counter-defense | Counter-defense |
| No defense | 3, 2 | 3, 1 | 3, 0 |
| Partial defense | 11.7, –7.7 | 3.6, –0.62 | 2.8, –0.8 |
| Defense | 10.9, –7.9 | 2.6, –0.66 | 1.9, –0.9 |

*The Nash equilibrium is indicated in blue.

It can be seen that the host has maximum payoff when it defends partially against the pathogen when the latter does not use counter-defense. Starting from any cell of Table 5, the equilibrium is reached at the cell 'Partial defense/ Partial counter-defense'. If the players deviate from these strategies, their payoff decreases. Thus, in this situation, no oscillation occurs, in contrast to the two-strategy game. Rather, a stationary trade-off is found.

It is of interest to analyze a three-strategy game where the payoffs are changing monotonically, in contrast to Table 5. For example, when 'Partial defense/ Partial counter-defense' provide payoff values of 2.8 and –4, there is an incentive for either player to leave the intermediate strategies. Then, a cyclic behavior as in Table 6 occurs. Note the important difference between monotonic and non-monotonic payoff functions. While the former lead to oscillations, the latter may lead to a pure Nash equilibrium.

### 3.3 Comparison to two-strategy game

To make the point of oscillation vs. stationary trade-off clearer, we simplified Table 5 by omitting the intermediate strategies, shown in Table 6. The payoff values for a two-strategy game in this table fulfil the same order relations as in the high-benefit case in Subsection 2.1.1 (Table1).

Thus, the game shows a cyclic dominance structure rather than a pure Nash equilibrium.

**Table 6:** Payoff values for the two-strategy host-pathogen game.*

| Host | Pathogen | |
|---|---|---|
| | No counter-defense | Counter-defense |
| No defense | 3, 2 ↓ | ← 3, 0 |
| Defense | 10.9, −7.9 → | ↑ 1.9, −0.9 |

*This payoff matrix leads to a mixed Nash equilibrium rather than a pure one. When following increases in payoff, a counter-clockwise cycle occurs, as indicated by arrows (see also Figure 7).

For the two-strategy game, a mixed Nash equilibrium can be calculated, meaning that each strategy is adopted by a certain probability. Let us consider that the host uses 'defense' strategy with probability $r$ and 'no defense' with probability $1-r$. This could be interpreted in that a fraction $r$ of all hosts use the defense strategy. Similarly, the pathogen uses 'counter-defense' strategy with probability $s$ and 'no counter-defense' with probability $1-s$.

Mixed Nash equilibria can be calculated by putting the derivative of the average payoffs with respect to probabilities equal to zero (as outlined in Subsection 2.1.1).

We obtain $r = \frac{2}{10.8} = 0.185$ and $s = \frac{7.9}{9} = 0.878$.

As soon as one player somewhat deviates from the best (mixed) response, the other player has an incentive to switch to a pure strategy, as can be shown by differentiating the above response functions. Hence the best response functions are as follows:

$$r(s) = \begin{pmatrix} 1, \text{if } s < 0.878, \\ 0, \text{if } s > 0.878, \\ (0,1), \text{if } s = 0.878 \end{pmatrix} \text{ and } s(r) = \begin{pmatrix} 1, \text{if } r > 0.185, \\ 0, \text{if } r < 0.185, \\ (0,1), \text{if } r = 0.185 \end{pmatrix}$$

For $s = 0.878$, the host can, as an immediate response, use any probability r. However, if this probability deviates from $r = 0.185$, the pathogen can, in the next iteration, respond accordingly and increase its payoff. Therefore, it is best for the host to choose $r = 0.185$ and analogously for the pathogen to choose $s = 0.878$, leading to the Nash equilibrium.

In terms of populations, the mixed Nash equilibria can be interpreted such that the probabilities correspond to fractions of the two populations. So, about 19% of the host population play the defense strategy and about 88% of the pathogen population play the counter-defense strategy.

In Table 7, the results of all games analyzed above are summarized, most of them in dependence on the order relation among benefits and costs. For simplicity's sake, for the three-strategy game, we did not distinguish any subcases depending on benefit and costs nor on perfectly and imperfectly efficient counter-defense. It is a general result that in the low-benefit case, always 'No defense/ No counter-defense' results. The explanation for the case "pathogen plays first" is given in the Supplement.

**Table 7:** Systematic overview of Nash equilibria for host-pathogen interactions in different cases.

| | Perfectly efficient counter-defense | | |
|---|---|---|---|
| | Non-sequential game | Sequential game | |
| | | Host plays first | Pathogen plays first |
| $b > c$ | No pure Nash equilibrium | No defense/ No counter-defense | Counter-defense/ No defense |
| $b < c$ | No defense/ No counter-defense | No defense/ No counter-defense | No counter-defense/ No defense |

| | Imperfectly efficient counter-defense | | |
|---|---|---|---|
| | Non-sequential game | Sequential game | |
| | | Host plays first | Pathogen plays first |
| $b/2 > c$ | Defense/ Counter-defense | Defense/ Counter-defense | Counter-defense/ defense |
| $b/2 < c < b$ | Defense/ No counter-defense | Defense/ No counter-defense | Counter-defense/ No defense |
| $b < c$ | No defense/ No counter-defense | No defense/ No counter-defense | No counter-defense/ No defense |

| Defense or counter-defense with three strategies |
|---|
| Non-monotonic payoff functions: Pure Nash equilibrium Partial defense/ Partial counter-defense |
| Monotonic payoff functions: Mixed Nash equilibrium |

# 4    Discussion

Here, we have analyzed, by a game-theoretical approach, the defense and counter-defense (for example, by chemical substances) among hosts and parasites. We have dealt with the paradox that when the host starts defending against the pathogen to increase its payoff, the pathogen is encouraged to counter-defend against the host to protect itself. If this counter-defense neutralizes the host's defense, the host can save the cost for it and switch it off. Then the pathogen can stop its counter-defense to save its cost (22). This may lead to an endless cycle of switching on and off the defense respectively counter-defense and corresponds to a mixed Nash equilibrium. Such equilibria were also found in a model describing the interaction between pathogenic bacteria and the human host, where the two strategies for either side correspond to the intracellular and extracellular locations (27) and between macrophages and fungal pathogens, where the two strategies for either side are 'aggressive' and 'peaceful' (31).

Games only having a mixed Nash equilibrium rather than a pure equilibrium include several asymmetric two-strategy games and also several three-strategy games, even if the latter are symmetric. A famous example is the rock-scissors-paper game (35, 46, 47). An example of a two-strategy game without a pure Nash equilibrium is provided by the 'matching pennies game' (35-37), as discussed in the Introduction. To our knowledge, our results are novel because the generalized matching pennies game is here used to explain chemical-ecological interactions. A related game is 'hard love' [(35), Section 6.6]: A mother supports her son financially if he seeks a job. However, he does not do so if he is supported and enjoys his leisure time instead. Therefore, mother stops the support, which prompts him to seek a job. Mother is happy and supports him again, and so on.

We determined the Nash equilibria for sequential and non-sequential host-pathogen interactions. First, we only considered two strategies for each player and used normal form games and game trees, as often done in game theory. For the sequential games, we considered both the cases that the pathogen plays first (see Supplement) and that the host plays first. Depending on the case considered, different Nash equilibria have been obtained. In the former case, the Nash equilibrium "No defense/Counter-defense" is obtained for certain parameter ranges. This is a counter-intuitive result because one may wonder why a counter-defense is beneficial if there is no defense. The reason is that the pathogen safeguards itself against both possible responses by the host. Although it may appear to be unrealistic that the pathogen chooses its strategy first, it can appropriately describe the interplay if the pathogen jumps from one host species to another one.

We analyzed the game depending on benefits and costs as well as on whether the counter-defense is perfect or imperfect. For example, in the case of imperfectly efficient enzymes and high benefit-to-cost ratios, the pure Nash equilibrium 'Defense/ Counter-defense' is obtained. Moreover, we found that host and pathogen can interact without any defense or counter-defense if the costs of producing the toxin and enzyme are higher than the benefit.

Thereafter, we considered a continuous spectrum of strategies for the host and pathogen. We simulated their payoffs by a Hill-type dose-response curve (24, 49, 50). Since we have to subtract the costs, which can be assumed to depend linearly on the dose, the payoff can show a maximum for the intermediate strategy for appropriate parameter values (24). A maximum can also occur in the simpler Simms-Rausher cost/benefit model, which is based on a Michaelis-Menten type function (51). Here, however, we use a Hill function because it describes the dose dependence more realistically. Then, we discretized the continuous spectrum of strategies by considering three possible actions and used normal form games. We obtained 'partial (counter)defense' strategies as the pure Nash equilibrium, which implies that permanently producing a moderate amount of toxin and enzyme rather than a cyclic switching on and off is the best choice for the organisms. To obtain this result, it is important to consider (at least) three strategies here.

The above-mentioned result that in certain parameter ranges, neither a defense nor a counter-defense occurs, is an idealization based on the binary discretization. The immune response is indeed often very costly so that a trade-off with the pathogen has to be reached (52). Earlier, it was shown by a game-theoretical model assuming a non-sequential game that pathogens are sometimes tolerated (53). However, such trade-offs usually include a certain (possibly low) extent of defense and counter-defense. This can be modeled adequately by the three-strategy game. The explanation of 'partial (counter)defense' is likely to have a very broad importance for host-pathogen, plant-herbivore and predator-prey interactions beyond chemical defense. Trade-offs (e.g. between resistance and tolerance) in which a submaximal effort is invested by either side are very relevant in evolution (2, 26, 52-54).

It is worth comparing that result with the outcome for the case where the dose-payoff curve (i.e. response minus cost) is monotonic. In the latter case, a strategy in an extreme situation (e.g. no defense or full defense) is chosen since a strategy with a higher payoff dominates a strategy with a lower payoff. The three-strategy game can then be simplified to a two-strategy game, which we have treated in the first part of the paper. This may lead to mixed Nash equilibria. Thus, the way the above-mentioned paradox is resolved depends on several conditions and parameters. In particular, it is relevant whether the counter-defense is very efficient, whether the game is sequential or non-sequential, whether the costs exceed the benefit and whether the dose-payoff curves are monotonic. The choice of payoff values may look quite arbitrary. However, payoff matrices can be scaled by adding a constant and by multiplying all payoffs by the same positive factor without changing the pure Nash equilibria. Thus, for these equilibria, only the order relations among the payoffs matter (35, 40).

Mixed Nash equilibria can have different biological implications. Either, a cyclic behavior occurs in that defense and no defense alternate on evolutionary time scales. Silent genes (pseudo-genes) may be an evolutionary remnant of another strategy used earlier (55, 56). As discussed above, another option is that both players only produce a certain percentage of the maximum amount of defense and counter-defense, respectively. It is interesting that a model using two discrete strategies for either player points, in the mixed Nash equilibrium, to a third possible strategy. A further possible interpretation is that some hosts use a (full) defense while others do not (and analogously some pathogens use or do not use a counter-defense) and they stay with their strategy. This might explain why some host species use a strong defense while other species do not.

For example, it is interesting that some bacteria, such as Salmonella species, produce ß-lactamases, while others, such as *Streptococcus pneumoniae*, do not (57). Our results suggest that the ß-lactamase-producing species evolved under conditions where such a counter-defense is an appropriate strategy, which corresponds to the Nash equilibrium Defense/ Counter-defense while the other bacteria evolved under conditions favoring the Nash equilibrium Defense/ No counter-defense. Another explanation is that the behavior of the different species corresponds to different points in the cycle of a mixed Nash equilibrium.

The analysis is also applicable to plant-herbivore or fungus-fungivore interactions (15-18). Our results show that depending on conditions, defense or no defense is the better option. This may explain why there are both toxic and non-toxic plants. However, it is worth noting that not every edible plant is non-toxic because the toxin concentration is often so low that it can be tolerated by the human body and even gives a spicy flavor. Examples are provided by many edible Brassicaceae such as cabbage, rape and mustard. It is also worth mentioning that for some plant species, different varieties differ in their production of toxins. For example, the subspecies *Cannabis sativa* produces higher amounts of -tetrahydrocannabinol than the subspecies *Cannabis indica* (58). The rowan plant (*Sorbus aucuparia*) produces parasorbic acid, which irritates the gastric mucosa in humans. A variety of this plant, called *Sorbus aucuparia var. moravica* has a lower amount of parasorbic acid and, thus, can be eaten raw (59).

## 5    Conclusion and future prospects

In our paper, we used the classical Nash approach to game theory, which leads to new interesting insights. Such an approach was used in analysing host-pathogen interactions earlier (27-29, 31, 32). Another approach is based on the concept of evolutionarily stable strategies, which allows one to study the fate of a rare mutant with a new strategy (40, 41). However, that concept in its basic form is only applicable to symmetric games, although extensions to asymmetric situations were proposed (35, 40). Importantly, every evolutionarily stable strategy corresponds to a Nash equilibrium but not necessarily vice versa. To check whether the mixed Nash equilibrium found above is evolutionarily stable, we refined the model to a three-strategy game and determined the payoffs using a dose-response curve.

To determine the stability of the obtained equilibria and to distinguish the different interpretations of the mixed Nash equilibrium, ordinary differential equations (ODEs) such as replicator equations (40), adaptive dynamics (60-62) or Lotka-Volterra equations (40, 46) are worth being used in future extensions of the model. Indeed, as an alternative to game theory, the interplay between defense and counter-defense is sometimes described quantitatively by ODEs (10, 25).

Both approaches have their pros and cons. The ODE approach can describe the time course but it needs a higher number of parameter values. It is very useful in distinguishing between oscillations in strategies and use of strategies with certain probabilities. For example, the time course in the rock-scissors-paper game can be simulated (46, 47). If the oscillations are damped, they tend to an asymptotically stable state. Then, the two interpretations of the mixed Nash equilibrium in terms of oscillations and in terms of (asymptotically) stationary frequencies of strategies in the population are in line with each other.

Game theory has the advantage of having a much wider scope because the concrete nature and details of interactions do not matter that much. Here, we used the situation of a toxin and an enzyme degrading that as the paradigm. However, the game-theoretical analysis is more general. It applies whenever an organism uses a perfectly or imperfectly counter-defense against the defense of another organism and the net payoffs can be quantified for both.

An interesting question is whether information exchange between the players is allowed. As outlined in the Introduction, we here use an interpretation in terms of populations, in which the equilibrium is found by natural selection without the necessity that the organisms have cognitive capabilities. Nevertheless, it will be interesting to shed light on the role of communication because many microorganisms can sense effector molecules produced by other organisms (42-44) and there is even cross-talk between pathogens and the human immune system (54).

The present analysis bears manifold potential applications in fighting pathogens. A major problem in clinical treatments of bacterial infections is the increasing resistance due to, for example, ß-lactamases. A possible extension to our model is to include the effect of counter-counter defenses (4, 10) into the payoff matrices.

A promising application of our study concerns the evolution of drug resistance. For example, *Plasmodium falciparum*, the causative agent of malaria, keeps evolving resistance to pharmaceuticals such as chloroquine (63). The two strategies of the patient would be to take or not to take the drug, while the two strategies of *P. falciparum* are to evolve or not to evolve resistance. Obviously, this is related to a matching-pennies game.

## 6      Author Contributions

StS conceptualized, coordinated and supervised the study. SD, RG and StS established the model. SD performed the calculations and produced the figures. SD and StS wrote the manuscript and all authors reviewed it.

## 7      Acknowledgments

Financial support by the Carl-Zeiss-Stiftung in the Jena School of Microbial Communication (to SD), the Max Planck Society through the IMPRS 'Exploration of Ecological Interactions with Molecular and Chemical Techniques' (to RG) and the German Research Foundation (DFG) through the TR 124 'Pathogenic fungi and their human host: Networks of Interaction', project number 210879364, is gratefully acknowledged.

The authors would like to thank Jan Ewald (Leipzig), Rosalind Allen and Suman Chakraborty (Jena), Christian Kost and Leonardo Oña (Osnabrück) as well as Gary Fogel (San Diego) for stimulating discussions.